\documentclass[apj]{emulateapj}
\usepackage{apjfonts}

\gdef\vdc{vDC}
\gdef\kms{km\,s$^{-1}$}
\gdef\msun{${\rm M}_{\odot}$}
\lefthead{van Dokkum \& Conroy}
\righthead{Dwarf-sensitive features in elliptical galaxies}
\slugcomment{Published in ApJ Letters}
\begin{document}

\title{Confirmation of enhanced dwarf-sensitive absorption features
in the spectra of massive elliptical galaxies: further evidence for
a non-universal initial mass function}

\author{Pieter G.\ van Dokkum\altaffilmark{1} and
Charlie Conroy\altaffilmark{2}
}

\altaffiltext{1}
{Department of Astronomy, Yale University, New Haven, CT 06520, USA}
\altaffiltext{2}{Harvard-Smithsonian Center for Astrophysics,
60 Garden Street, Cambridge, MA 02138, USA}

\begin{abstract}

  We recently found that massive cluster elliptical galaxies have
  strong Na\,I $\lambda 8183,8195$ and FeH $\lambda 9916$ Wing-Ford
  band absorption, indicating the presence of a very large population
  of stars with masses $\lesssim 0.3$\,\msun. Here we test this result
  by comparing the elliptical galaxy spectra to those of luminous
  globular clusters associated with M31. These globular clusters have
  similar metallicities, abundance ratios and ages as massive
  elliptical galaxies but their low dynamical mass-to-light ratios
  rule out steep stellar initial mass functions (IMFs). From high
  quality Keck spectra we find that the dwarf-sensitive absorption
  lines in globular clusters are significantly weaker than in
  elliptical galaxies, and consistent with normal IMFs.  The
  differences in the Na\,I and Wing-Ford indices are $0.027\pm 0.007$
  mag and $0.017\pm 0.006$ mag respectively.  We directly compare the
  two classes of objects by subtracting the averaged globular cluster
  spectrum from the averaged elliptical galaxy spectrum.  The
  difference spectrum is well fit by the difference between a stellar
  population synthesis model with a bottom-heavy IMF and one with a
  bottom-light IMF.  We speculate that the slope of the IMF may vary
  with velocity dispersion, although it is not yet clear what physical
  mechanism would be responsible for such a relation.

\end{abstract}


\section{Introduction}

The form of the stellar initial mass function (IMF) is one of the main
uncertainties in the interpretation of observations of distant
galaxies and in our understanding of the conversion of gas to stars
over cosmic time.  Most of the stellar mass density of the Universe is
in the form of low mass stars with masses $\ll 1$\,\msun. Despite the
large contribution of these stars to the total stellar mass of
galaxies their contribution to the integrated light is small,
which makes it difficult to measure the form of the IMF in galaxies
other than the Milky Way. In practise it is usually
assumed that the IMF
in other galaxies and at earlier epochs is the same as in the disk of
the Milky Way: a power-law with logarithmic slope $x=-2.3$ at masses
$\gtrsim 1$\,\msun with a turnover at lower masses ({Kroupa} 2001; {Chabrier} 2003; {Bastian}, {Covey}, \& {Meyer} 2010).

As has long been recognized it is possible to measure the contribution
of dwarf stars to the integrated light of old stellar populations from
absorption features that have a strong gravity dependence
(e.g., {Spinrad} 1962; {Cohen} 1978; {Faber} \& {French} 1980). There are
two absorption features in the red that are strong in cool dwarfs but
very weak or absent in cool giants: the Na\,I $\lambda
8183,8195$\,\AA\ doublet (e.g., {Schiavon} {et~al.} 1997) and the
Wing-Ford FeH molecular band at $\lambda 9916$\,\AA\
(e.g., {Wing} \& {Ford} 1969; {Schiavon}, {Barbuy}, \& {Bruzual  A.} 2000). These features reach depths of
30-40\,\% and 40--50\,\% respectively in the spectra of individual low
mass stars. In the integrated light of old stellar populations their
predicted strengths are of order 1--5\,\% depending on the form of the
IMF (see {van Dokkum} \& {Conroy} 2010, herafter \vdc).

In \vdc\ we presented high quality Keck spectra for eight massive
galaxies in the Coma and Virgo clusters.  Both the Na\,I doublet and
the Wing-Ford band are quite strong in these spectra, in agreement
with previous studies (e.g., {Carter}, {Visvanathan}, \&  {Pickles} 1986; {Hardy} \& {Couture} 1988; {Couture} \& {Hardy} 1993).
As shown in \vdc, the absorption is
significantly stronger (by a factor of $\sim 3$) than might be
expected from a 10\,Gyr old Solar metallicity stellar population with
a Milky-Way IMF. The best-fitting IMF is more dwarf-enriched than
even the {Salpeter} (1955) form, requiring a slope $x\sim -3$ down
to 0.1\,\msun (compared to $x=-2.35$ for a Salpeter IMF). 

The main uncertainty in \vdc\ is that we rely on stellar population
synthesis models. The models are based on high-quality empirical
spectra of individual Solar metallicity stars from the
{Rayner}, {Cushing}, \& {Vacca} (2009) IRTF library of cool stars, and use up-to-date
isochrones (see {Conroy}, {Gunn}, \& {White} 2009, vDC, and C.~Conroy et al., in
preparation).
However, it is known that the stars in elliptical galaxies are
enhanced in $\alpha$-elements compared to stars in the Milky Way
({Worthey}, {Faber}, \&  {Gonzalez} 1992) and
this cannot be incorporated explicitly in our modeling.


In this {\em Letter} we 
test whether the enhancement of
dwarf-sensitive spectral features in elliptical galaxies persists when
we compare their spectra to those of globular clusters rather than to
models. The Andromeda galaxy (M31) has a population
of old, metal-rich globular that have
similar abundance patterns as massive
elliptical galaxies ({Caldwell} {et~al.} 2011).  However, their dynamical
mass-to-light ($M/L$) ratios are so low that they must have normal
mass functions -- or possibly even mass functions that are
dwarf-deficient (``bottom-light'') compared to the Milky Way IMF
({Strader} {et~al.} 2011). Therefore, if the interpretation in
\vdc\ is correct, we should see much weaker Na\,I and FeH absorption
than in massive elliptical galaxies, despite the fact that their ages
and abundances are very similar.

\section{Sample selection and observations}

\begin{figure}[t]
\epsfxsize=8cm
\epsffile{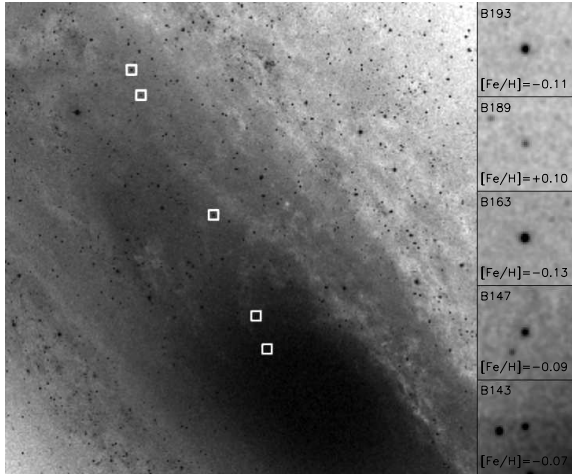}
\caption{\small The five globular clusters that were observed with
LRIS. The main image is an optical image
from the Digital Sky Survey and the
insets are $J$ band images from 2MASS. Four of the clusters
are among the brightest in M31 and have approximately
Solar metallicity (Caldwell et al.\ 2011).
The relatively faint cluster B189 was added to the sample to test
how sensitive the conclusions are to abundance patterns.
\label{m31.plot}}
\end{figure}

\begin{figure}[!t]
\epsfxsize=8cm
\epsffile{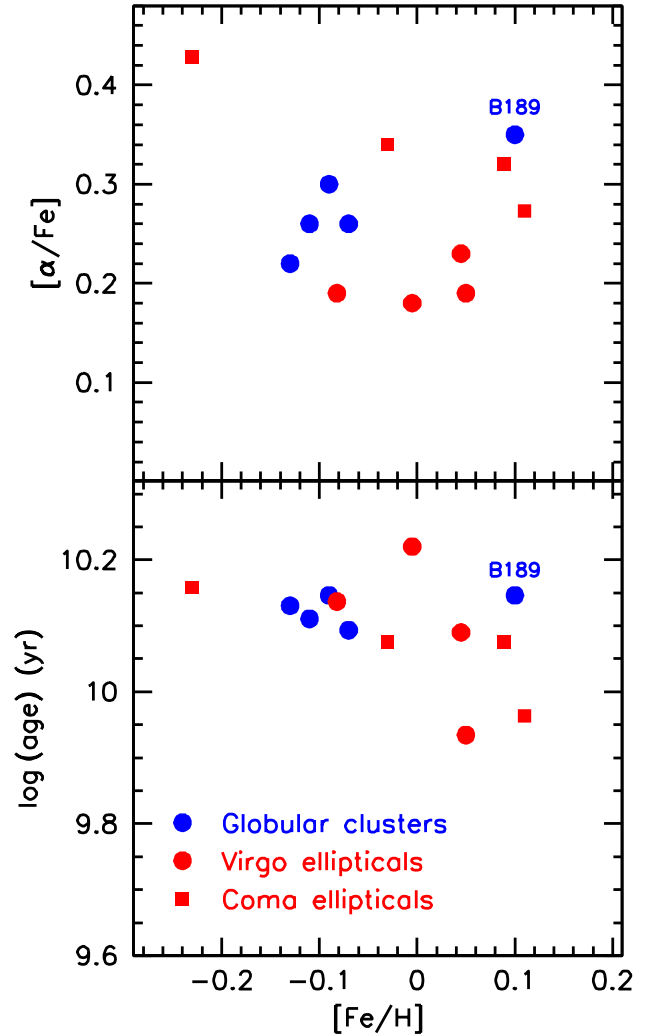}
\caption{\small Comparison of the iron abundances [Fe/H],
$\alpha$-enhancement [$\alpha$/Fe], and ages of the
M31 globular clusters to those of the elliptical galaxies
of vDC. The M31 clusters have similar
abundance patterns and ages as the elliptical galaxies.
Typical errors in individual measurements are $0.1-0.2$\,dex, i.e.,
most of the data points are within 1$\sigma$ of one another. 
\label{select.plot}}
\end{figure}

We observed four bright globular clusters associated
with M31 for which ages and metallicities had been
determined in other studies:
B143, B147, B163, and B193.  The clusters were selected to
be bright and old and to have
approximately Solar iron abundances
(Caldwell et~al.\ 2011).
For a Chabrier (2003) IMF the total
number of stars in these four clusters is
approximately $10^7$ (of which $\sim 100$ should be AGB stars),
which is
sufficient to properly sample the stellar luminosity function. We also
observed the very metal-rich cluster B189 to assess
the sensitivity of the FeH and Na\,I features to changes in
abundance patterns (see below).  The
locations of the clusters within M31 are indicated in Fig.\
\ref{m31.plot}.  In Fig.\ \ref{select.plot} their abundances and ages
are compared to those of the elliptical galaxies of \vdc.  The
globular cluster data come from Caldwell et al.\ (2011) and R.\
Schiavon et al., in prep. Data for the four Virgo galaxies were
obtained from {Trager} {et~al.} (2000), for an $r_e/8$ aperture (which is
similar to the effective aperture of our measurements of Virgo
galaxies). The Coma measurements are from {Thomas} {et~al.} (2005) (NGC\,4840
and NGC\,4926) and from {Harrison} {et~al.} (2010) (NGC\,4889 and IC\,3976).
All measurements should be on approximately the same system, although
systematic differences between studies cannot be ruled out.
The four luminous M31 globular clusters have similar ages and
abundances as the elliptical galaxies.  The cluster B189 has a higher
iron abundance and $\alpha-$enhancement than most of the ellipticals.
Furthermore, it is extremely enhanced in C and N,
and it has strong NaD absorption (R.\ Schiavon, priv.\ comm.).
\noindent

The globular clusters were observed on December 3, 2010 with the red arm of
the Low Resolution and Imaging Spectrometer (LRIS; {Oke} {et~al.} 1995) on
the Keck I telescope.  We used the 600\,l\,mm$^{-1}$ grating blazed
at 1\,$\mu$m. The slit width of $0\farcs 7$ gives a spectral
resolution of $\sigma = 1.2$\,\AA\ at 9500\,\AA, corresponding to a
velocity resolution of $\approx 40$\,\kms. Each cluster was dithered
along the slit in a series of four 300\,s exposures.  Data reduction
followed standard procedures for long-slit spectroscopy.
Owing to the fully-depleted
LBNL CCDs of LRIS fringing is not a concern. 
We corrected for atmospheric
absorption in the following way. Before and after each globular
cluster we observed the Galactic A star HIP\,2860.  Next, we convolved
a high S/N Mauna Kea night sky spectrum
to the instrumental
resolution and scaled the absorption to match strong
observed atmospheric absorption lines in
HIP\,2860. Finally, for each globular cluster we interpolated the two
scaled absorption spectra obtained from the HIP\,2860 observations
bracketing that cluster.  This procedure worked well, leaving no
detectable residuals of night sky absorption lines in the wavelength
regions of interest.
\noindent
\noindent
\begin{figure*}[!t]
\epsfxsize=17.5cm
\epsffile{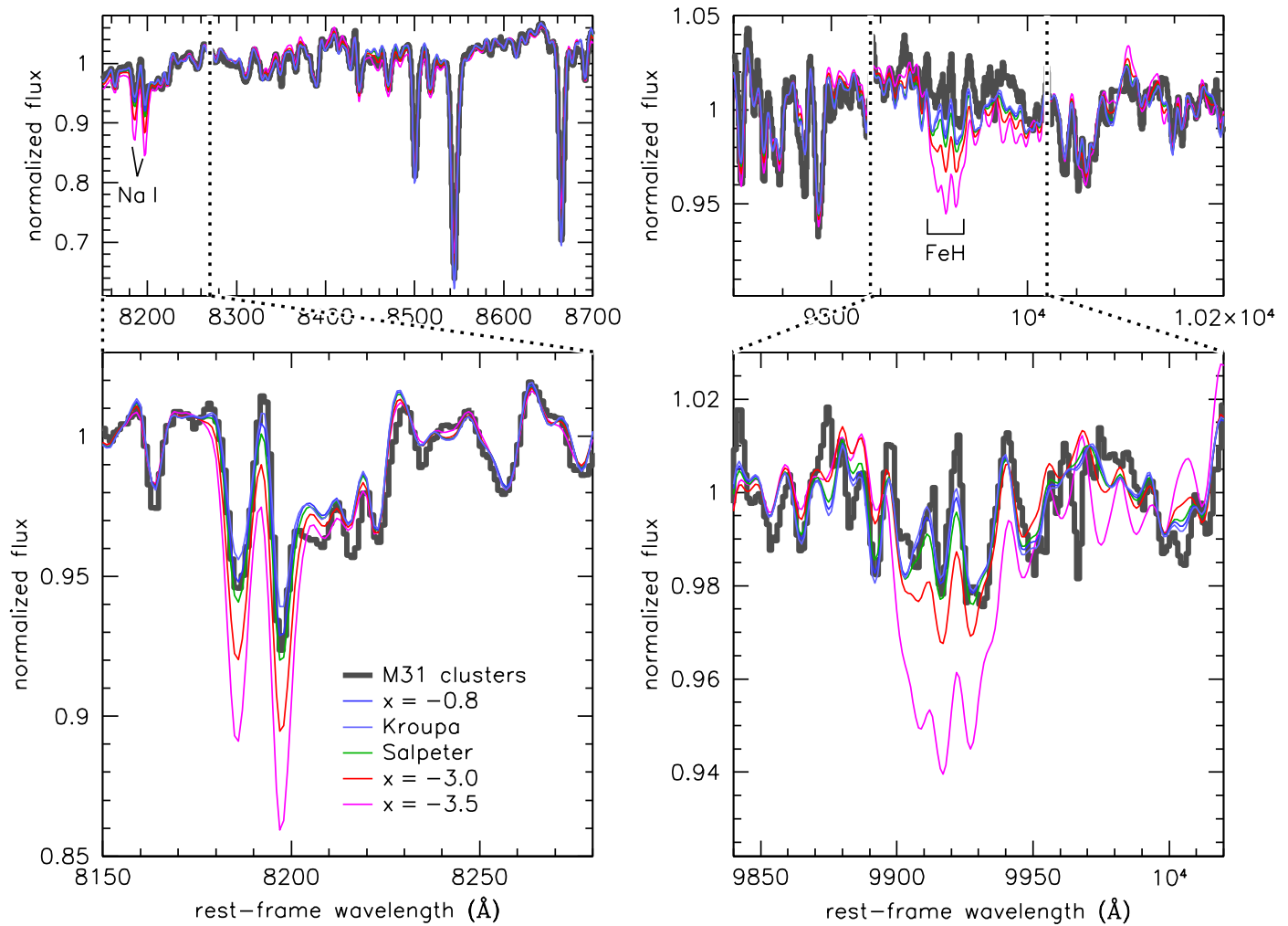}
\caption{\small Averaged Keck spectra of the M31 globular clusters
B143, B147, B163, and B193
(thick grey lines). The left panels show the region near the
Na\,I doublet and the right panels show the FeH Wing-Ford band
region. The bottom panels zoom in on the IMF sensitive features.
Colored lines show predictions from stellar population
synthesis models with different IMFs.
Models with ``normal'' IMFs match the data very well over the
entire wavelength range. Dwarf-rich IMFs with steep slopes
($x\sim -3$) --- which provide good fits to massive
elliptical galaxies --- do not fit the observed
weak Na\,I doublet and
Wing-Ford band of the globular clusters.
\label{globs.plot}}
\end{figure*}

One-dimensional spectra were extracted by summing the central 13
rows of the two-dimensional spectra, corresponding to $1\farcs8$
(7 pc, which is approximately the half-light diameter of the clusters).
The results are independent of the extraction aperture; as expected
(see \S\,5) we are not able to detect mass segregation in these clusters.
The spectra of the four most luminous globular clusters were
de-blueshifted to restframe wavelengths, normalized, and averaged in
order to increase the S/N ratio and diminish the effects of any
remaining systematic detector or night sky residuals. The averaging of
these four clusters is justified as their ages and metallicities are
nearly identical (see Fig.\ \ref{select.plot}).  In the following the
cluster B189 will be discussed separately, although
we note here that including this cluster in the average spectrum has
no impact on our analysis or conclusions.

\section{The Na\,I doublet and the Wing-Ford band in M31
Globular Clusters}

Figure\ \ref{globs.plot} shows the averaged spectrum of the M31
globular clusters in the wavelength regions near the Na\,I doublet
(left panel) and the Wing-Ford band (right panel). Note that the Na\,I
doublet is resolved and not blended with TiO, as globular clusters
have much lower velocity dispersions than the elliptical galaxies
discussed in \vdc.  Following \vdc\ the spectrum was normalized by
fitting second-order polynomials in the top panels (excluding the
regions around the features of interest) and first-order polynomials
in the bottom panels.  The typical uncertainty in the averaged
spectrum is 0.003\,\AA$^{-1}$, as judged from the median scatter among
the four globular clusters over the wavelength range 8200\,\AA\ --
8400\,\AA.  All the easily visible absorption lines are stellar
features, not noise. We smoothed the spectra slightly to approximate
the resolution of the IRTF spectral library (see below).

Colored lines are stellar population synthesis models. The models are
the same as those in \vdc: they are 10 Gyr old, Solar
metallicity models based on empirical stellar spectra from the
{Rayner} {et~al.} (2009) IRTF library. Models with slightly different
ages are virtually identical. The models will be presented in
detail in C.~Conroy et al., in preparation.
Different colors indicate different
dwarf contributions, ranging from bottom-light to bottom-heavy
IMFs. The data are well fit by models with a {Kroupa} (2001) or
bottom-light IMF: the median absolute deviation is 0.005\,\AA$^{-1}$
from 8150\,\AA\ -- 8700\,\AA\ and $0.007$\,\AA$^{-1}$ from 9700\,\AA\
-- 10,200\,\AA. This demonstrates that our stellar population
synthesis model, which is based on Solar metallicity stars in the
Milky Way, provides an adequate description of the spectra of globular
clusters in M31.

We have shown in \vdc\ that elliptical galaxies are best fit by a
bottom-heavy IMF with $x=-3$ (the red line in Fig.\
\ref{globs.plot}). The globular clusters are poorly fit by such
models. In the Na\,I band (8185\,\AA\ -- 9205\,\AA; see vDC) the
median absolute deviation is $0.025\pm 0.005$ for the $x=-3$ IMF,
compared to 0.004 for a {Kroupa} (2001) IMF. In the Wing-Ford band
(9910\,\AA\ -- 9930\,\AA) the median absolute deviation is $0.015\pm
0.005$ for $x=-3$ and 0.003 for Kroupa.

\noindent
\begin{figure*}[htbp]
\epsfxsize=17.5cm
\epsffile{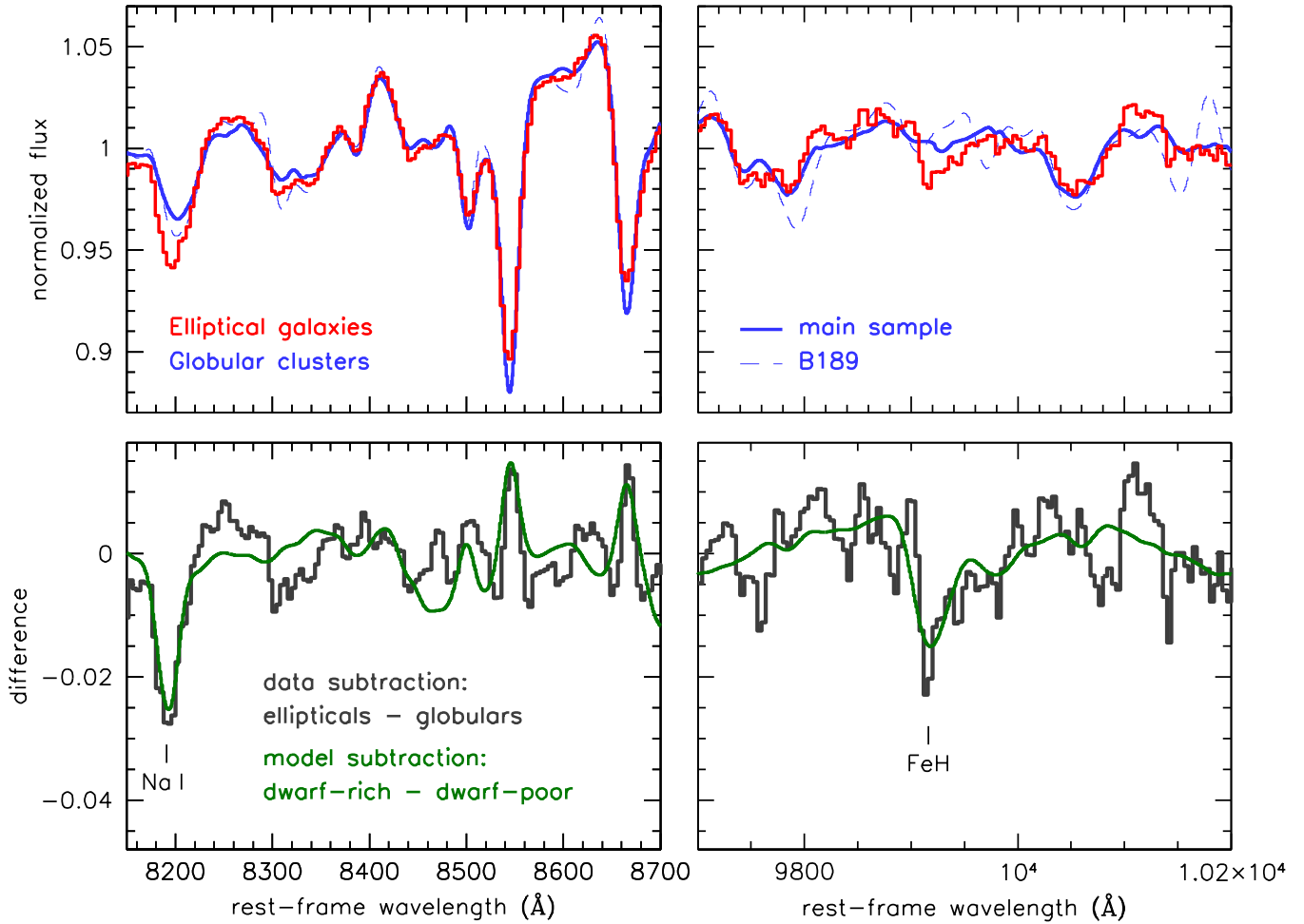}
\caption{\small
{\em Top panels:} Direct comparison of the globular cluster spectra
to the elliptical galaxy spectra.
The globular cluster
spectra were smoothed to match the average velocity dispersion
of the elliptical galaxies. The main sample comprises the luminous
clusters B143, B147, B163 and B193. The cluster B189
is shown separately; note that the S/N ratio of this spectrum is much lower
than that of the averaged spectrum of the other clusters.
{\em Bottom panels:} The black spectrum is the difference
between the elliptical galaxies
and the main sample of globular clusters. The Doppler-broadened
Na\,I doublet and the Wing-Ford band are clearly much stronger
in the elliptical galaxies than in the globular clusters.
The green line is the difference between a stellar population
synthesis model with an enhanced dwarf population ($x=-3$)
and one with a depressed dwarf population ($x=-0.8$).
The green line is a good fit to the black
line, which shows that IMF variations can explain the spectral
differences between globular clusters and massive elliptical galaxies.
\label{emp.plot}}
\end{figure*}

\section{Direct Comparison between globular clusters and elliptical
  galaxies}

The fact that stellar population synthesis models indicate a normal
IMF for globular clusters is significant as those same models
indicated a bottom-heavy IMF for elliptical galaxies.  We can also
compare the globular clusters to the elliptical galaxies in a more
direct way.  In the top panels of Fig.\ \ref{emp.plot} we show the
averaged cluster spectrum along with the averaged elliptical galaxy
spectrum.  The Coma and Virgo spectra of \vdc\ were averaged, and the
globular cluster spectrum was smoothed to the same
velocity width as the elliptical galaxies. As expected,
the averaged globular
cluster spectrum is very similar to the averaged
elliptical galaxy spectrum,
except in the regions around Na\,I and the Wing-Ford band.
The 
globular cluster B189 is shown separately. Although the absorption
lines in this cluster are slightly stronger than in the other clusters
owing to its higher metallicity, the Na\,I and Wing-Ford band do
not reach the depth seen in the ellipticals. This demonstrates that
our conclusions are insensitive to residual metallicity or
abundance variations
between globular clusters and elliptical galaxies.

The bottom panels of Fig.\ \ref{emp.plot} show the difference
spectrum, obtained by subtracting the averaged spectrum of the
globular clusters from that of the elliptical galaxies. The two most
significant features in the difference spectrum are at $\sim
8190$\,\AA\ and at $\sim 9920$\,\AA.  In both cases the absorption in
the ellipticals is much stronger than in the globular clusters. The
wavelengths of these features correspond to those of the
Doppler-broadened Na\,I doublet and the Wing-Ford band respectively,
demonstrating empirically that these dwarf-sensitive features are
significantly stronger in massive elliptical galaxies than in M31
globular clusters.  Expressed as an index difference, the Na\,I index
(as defined in \vdc) is $0.027 \pm 0.007$ mag stronger and the
Wing-Ford index is $0.017 \pm 0.006$ mag stronger, where the
uncertainties are determined from the scatter among the individual
objects entering the two spectra.

The green line in Fig.\ \ref{emp.plot} shows the difference between
two of the models shown in Fig.\ \ref{globs.plot}: a model with a
bottom-heavy IMF ($x=-3$, which was the best fit to the elliptical
galaxies in \vdc) and a model with a bottom-light IMF ($x=-0.8$, which
provides a good fit to the $M/L$ ratios of globular clusters
[Strader et al.\ 2011]).
The models have identical age (10 Gyr) and Solar metallicity;
the only difference is the number of low mass stars. The green line is
a good fit to the difference spectrum, as might have been expected
from the analyses in \vdc\ and \S\,3.  The model also reproduces the
fact that the Ca II $\lambda 8498$, $\lambda 8542$, and $\lambda 8662$
lines are {\em weaker} in the globular clusters; in the stellar
population synthesis model this reflects the weak Ca II absorption in
dwarfs compared to giants. Although giants dominate the light, for an
IMF as steep as $x=-3$ there are sufficient dwarfs to slightly depress
the Ca lines. The relative weakness of the Ca triplet in elliptical
galaxies was previously discussed by, e.g.,
{Cenarro} {et~al.} (2003).  We note, however,
that in our differential analysis this effect is very sensitive
to small errors in matching the velocity dispersions of the two
spectra.

\section{Discussion}

We have shown that massive cluster elliptical galaxies have much
stronger Na\,I and FeH absorption than old metal-rich globular
clusters in M31. This constitutes a critical test of the analysis in
\vdc: equally strong absorption in the globular clusters would have
implied that the enhancement in elliptical galaxies had incorrectly
been interpreted as an IMF effect, as the low $M/L$ ratios of
globular clusters are inconsistent
with dwarf-rich IMFs. Although we cannot exclude that
some unknown stellar population effect is responsible for the strong
absorption in elliptical galaxies, it is striking how well the
difference spectrum in Fig.\ \ref{emp.plot} is fit by an enhanced
dwarf population. Any explanation not involving dwarf stars has to
account for the enhanced Na\,I and FeH absorption in ellipticals, the
suppressed Ca II, and the similarity of TiO, Fe, Mg, H$\beta$ and
other elements. We note that known abundance anomalies in Galactic globular
clusters work in the opposite direction ({Gratton}, {Sneden}, \& 
{Carretta} 2004).

If dwarfs are indeed the cause of the differences in absorption then
Fig.\ 4 constitutes direct evidence for significant variation
in the low mass IMF. Quantifying the IMF in elliptical
galaxies using
stellar population synthesis models, we
can confidently rule out dwarf-suppressed IMFs such as those
proposed by {van Dokkum} (2008) and {Dav{\'e}} (2008) and also
``standard'' {Kroupa} (2001) or {Chabrier} (2003)
IMFs which are appropriate for the Milky Way disk. This conclusion
does not depend on the detailed present-day stellar mass function
of globular clusters (which may have been modified by dynamical
effects), as the expected strengths of the Na\,I line and
the Wing-Ford band are almost identical for all ``light'' IMFs
(Schiavon et al.\ 2000; Fig.\ 3).

Star formation likely proceeded
very differently in massive elliptical galaxies than in
the disks of spiral galaxies.
It is now thought that the progenitors of massive ellipticals
were very compact, with average densities $\gtrsim
20$\,\msun\,pc$^{-3}$ inside the effective radius
(e.g., {van Dokkum} {et~al.} 2008; {Buitrago} {et~al.} 2008). These densities are similar
to giant molecular clouds in the Milky Way, but given the $\sim$\,kpc
scale of these galaxies the column densities would have been several
orders of magnitude higher.  It will be interesting to measure the
physical conditions in the star-forming progenitors of these galaxies
and compare them to measurements in star-forming regions in the Milky
Way (see, e.g., {Ivison} {et~al.} 2010).

This study can be extended in many ways. The comparison between
globular clusters and elliptical galaxies itself is model-independent,
but we still rely on stellar population synthesis models to quantify
how steep the IMF is in elliptical galaxies. The main uncertainty
in these models is whether short-lived
stellar evolution phases with unusual abundance patterns are missed,
and this can be addressed by augmenting the spectral library.
Radial gradients in Na\,I and
FeH can provide information on the spatial distribution of the
dwarfs, which in turn might reflect different formation
mechanisms for
the central of ellipticals and their outskirts
 (e.g., {Oser} {et~al.} 2010; {van Dokkum} {et~al.} 2010).
Some studies find steep gradients in Na\,I ({Boroson} \& {Thompson} 1991),
which might suggest that such effects could be important.  It will
also be interesting to extend this study to elliptical galaxies of
lower luminosity. There is good evidence that the dynamical $M/L$
ratio of early-type systems scales with velocity dispersion
(e.g., {Cappellari} {et~al.} 2006; {Forbes} {et~al.} 2008; {Treu} {et~al.} 2010). A varying IMF
may be responsible for this trend ({Treu} {et~al.} 2010,
Dutton et al.\ 2010), although
some studies consider dark matter variations a more likely possibility
({Graves} \& {Faber} 2010).  As
noted by {Cappellari} {et~al.} (2006) the dynamical $M/L$ ratios of low
luminosity ellipticals may {\em require} a normal IMF, as they appear
to be lower than the $M/L$ ratio of a stellar population with a
bottom-heavy IMF ($M/L_B \approx 23$ and $M/L_R \approx 13$ for
$x=-3$). A preliminary analysis of a Keck spectrum of the elliptical
galaxy NGC~4458 ($\sigma = 85$\,\kms; {Cappellari} {et~al.} 2006)
shows that its dwarf-sensitive features are indeed similar to those
of the M31 globular clusters. Finally,
it will be interesting to re-examine studies of the evolution
of scaling relations of early-type galaxies 
(e.g., {Renzini} 2006). 
The color and $M/L$ evolution of massive galaxies
in clusters seem to require a bottom-light IMF with
$x=-0.7^{+0.7}_{-0.3}$ ({van Dokkum} 2008),
but it may be possible to fit the data with
steeper mass functions when structural evolution is incorporated
in the modeling
(e.g., {van der Wel} {et~al.} 2008; {van Dokkum} 2008; {Hopkins} {et~al.} 2009; {Holden} {et~al.} 2010).

A varying IMF has important implications for the interpretation of
observations of distant galaxies. The $M/L$ ratios of stellar
populations with an $x=-3$ IMF and a Kroupa IMF differ by
approximately a factor of 3--4.  If it is not known how the IMF
varies with galaxy type, this uncertainty directly corresponds to
the systematic uncertainty in the stellar masses and star formation
rates of galaxies. Even
though massive elliptical galaxies make up only a small fraction of
the galaxy population today, their old ages imply that their 
progenitors become increasingly more dominant at higher
redshifts.
As a result, the star formation history of the Universe,
the evolution of the galaxy mass function, and many other observations
would have to be revised. 

\acknowledgements{We are grateful to Ricardo Schiavon and Nelson
  Caldwell for providing metallicities and ages of M31 globular
  clusters prior to publication. We thank Judy Cohen, Richard Larson,
Ricardo Schiavon, and Scott Trager for useful discussions.}


\begin{references}

\reference{} {Bastian}, N., {Covey}, K.~R., \& {Meyer}, M.~R. 2010, \araa, 48, 339

\reference{} {Boroson}, T.~A. \& {Thompson}, I.~B. 1991, \aj, 101, 111

\reference{} {Buitrago}, F., {Trujillo}, I., {Conselice}, C.~J., {Bouwens}, R.~J.,  {Dickinson}, M., \& {Yan}, H. 2008, \apjl, 687, L61

\reference{} {Caldwell}, N., {Schiavon}, R., {Morrison}, H., {Rose}, J.~A., \& {Harding}, P.  2011, \aj, 141, 61

\reference{} {Cappellari}, M., {Bacon}, R., {Bureau}, M., {Damen}, M.~C., {Davies}, R.~L.,  {de Zeeuw}, P.~T., {Emsellem}, E., {Falc{\'o}n-Barroso}, J., {et al.} 2006, \mnras, 366, 1126

\reference{} {Carter}, D., {Visvanathan}, N., \& {Pickles}, A.~J. 1986, \apj, 311, 637

\reference{} {Cenarro}, A.~J., {Gorgas}, J., {Vazdekis}, A., {Cardiel}, N., \& {Peletier},  R.~F. 2003, \mnras, 339, L12

\reference{} {Chabrier}, G. 2003, \pasp, 115, 763

\reference{} {Cohen}, J.~G. 1978, \apj, 221, 788

\reference{} {Conroy}, C., {Gunn}, J.~E., \& {White}, M. 2009, \apj, 699, 486

\reference{} {Couture}, J. \& {Hardy}, E. 1993, \apj, 406, 142

\reference{} {Dav{\'e}}, R. 2008, \mnras, 385, 147

\reference{} {Dutton}, A.~A., et al.\ 2010, MNRAS, submitted (arXiv:1012.5859)

\reference{} {Faber}, S.~M. \& {French}, H.~B. 1980, \apj, 235, 405

\reference{} {Forbes}, D.~A., {Lasky}, P., {Graham}, A.~W., \& {Spitler}, L. 2008, \mnras,  389, 1924

\reference{} {Gratton}, R., {Sneden}, C., \& {Carretta}, E. 2004, \araa, 42, 385

\reference{} {Graves}, G.~J. \& {Faber}, S.~M. 2010, \apj, 717, 803

\reference{} {Hardy}, E. \& {Couture}, J. 1988, \apjl, 325, L29

\reference{} {Harrison}, C.~D., {Colless}, M., {Kuntschner}, H., {Couch}, W.~J., {de  Propris}, R., \& {Pracy}, M.~B. 2010, \mnras, 409, 1455

\reference{} {Holden}, B.~P., {van der Wel}, A., {Kelson}, D.~D., {Franx}, M., \&  {Illingworth}, G.~D. 2010, \apj, 724, 714

\reference{} {Hopkins}, P.~F., {Hernquist}, L., {Cox}, T.~J., {Keres}, D., \& {Wuyts}, S.  2009, \apj, 691, 1424

\reference{} {Ivison}, R.~J., {Swinbank}, A.~M., {Swinyard}, B., {Smail}, I., {Pearson},  C.~P., {Rigopoulou}, D., {Polehampton}, E., {Baluteau}, J., {et al.} 2010, \aap, 518, L35

\reference{} {Kroupa}, P. 2001, \mnras, 322, 231

\reference{} {Oke}, J.~B., et al. 1995, \pasp, 107, 375

\reference{} {Oser}, L., {Ostriker}, J.~P., {Naab}, T., {Johansson}, P.~H., \& {Burkert}, A.  2010, \apj, 725, 2312

\reference{} {Rayner}, J.~T., {Cushing}, M.~C., \& {Vacca}, W.~D. 2009, \apjs, 185, 289

\reference{} {Renzini}, A. 2006, \araa, 44, 141

\reference{} {Salpeter}, E.~E. 1955, \apj, 121, 161

\reference{} {Schiavon}, R.~P., {Barbuy}, B., \& {Bruzual A.}, G. 2000, \apj, 532, 453

\reference{} {Schiavon}, R.~P., {Barbuy}, B., {Rossi}, S.~C.~F., \& {Milone}, A. 1997, \apj,  479, 902

\reference{} {Spinrad}, H. 1962, \apj, 135, 715

\reference{} {Strader}, J., {Caldwell}, N., \& {Seth}, A. 2011,
AJ, in press (arXiv:1104.4649)

\reference{} {Thomas}, D., {Maraston}, C., {Bender}, R., \& {Mendes de Oliveira}, C. 2005,  \apj, 621, 673

\reference{} {Trager}, S.~C., {Faber}, S.~M., {Worthey}, G., \& {Gonz{\'a}lez}, J.~J. 2000,  \aj, 119, 1645

\reference{} {Treu}, T., {Auger}, M.~W., {Koopmans}, L.~V.~E., {Gavazzi}, R., {Marshall},  P.~J., \& {Bolton}, A.~S. 2010, \apj, 709, 1195

\reference{} {van der Wel}, A., {Holden}, B.~P., {Zirm}, A.~W., {Franx}, M., {Rettura}, A.,  {Illingworth}, G.~D., \& {Ford}, H.~C. 2008, \apj, 688, 48

\reference{} {van Dokkum}, P.~G. 2008, \apj, 674, 29

\reference{} {van Dokkum}, P.~G. \& {Conroy}, C. 2010, \nat, 468, 940 [vDC]

\reference{} {van Dokkum}, P.~G., {Franx}, M., {Kriek}, M., {Holden}, B., {Illingworth},  G.~D., {Magee}, D., {Bouwens}, R., {Marchesini}, D., {et al.} 2008, \apjl, 677, L5

\reference{} {van Dokkum}, P.~G., {Whitaker}, K.~E., {Brammer}, G., {Franx}, M., {Kriek},  M., {Labb{\'e}}, I., {Marchesini}, D., {Quadri}, R., {et al.} 2010, \apj, 709, 1018

\reference{} {Wing}, R.~F. \& {Ford}, Jr., W.~K. 1969, \pasp, 81, 527

\reference{} {Worthey}, G., {Faber}, S.~M., \& {Gonzalez}, J.~J. 1992, \apj, 398, 69

\end{references}




\end{document}